\documentclass[11pt,dvips]{article}
\usepackage{a4}
\usepackage{cite}
\usepackage{epsfig}
\usepackage{amssymb,amsmath}

\usepackage{sidecap}
\usepackage{caption2}

\newcommand{\tablecaption}{%
\setlength{\abovecaptionskip}{0pt}
\setlength{\belowcaptionskip}{10pt}
\caption}

\newcounter{pdfadd}
\usepackage[hyperindex,bookmarks,breaklinks]{hyperref}
\hypersetup{%
  pdftitle    = {On the Investigation of the Proton Structure Using QED 
                 Compton Events in ep Scattering},
  pdfauthor   = {V. Lendermann, H.-C. Schultz-Coulon, D. Wegener},
  pdfkeywords = {QEDC, DICS, proton structure function},
  pdfsubject  = { },
  pdfcreator  = {LaTeX with hyperref package + dvips + ps2pdf},
  pdfproducer = { }
}

\begin{document}

\title{\bfseries
On the Investigation of the Proton Structure\\
Using QED Compton Events in {\itshape ep} Scattering}

\author
{\Large \sc
V.\,Lendermann%
\footnote{Deutsches Elektronen-Synchrotron (DESY), D-22607 Hamburg, Germany.
\mbox{E-mail: victor@mail.desy.de}} ,
H.-C.\,Schultz-Coulon%
\footnote{Institut f\"ur Physik, Universit\"at Dortmund, D-44221 Dortmund, Germany} ,
D.\,Wegener%
\footnotemark[2]}

\date{}

\maketitle

\begin{abstract}
QED Compton scattering at HERA is discussed in terms of the information it may
reveal on the proton structure at low momentum transfers $Q^2$. Detailed Monte
Carlo studies are performed which show that the analysis of inelastic QED
Compton events allows the extension of present HERA structure function
measurements to a kinematic regime, which up to now was only accessed in fixed
target data. For these studies an improved version of the COMPTON event
generator is used, where special emphasis has been put on modelling the
hadronic final state at low invariant masses.

As the low $Q^2$ regime is sometimes also discussed in the context of the
collinear approximation and the possibility of measuring the photonic content
of the proton, the Monte Carlo studies are also used to check the validity of
this approach. It is found that the proposed concept of a photon density
$\gamma$ of the proton does not provide sufficient accuracy for the description
of inelastic QED Compton scattering.
\end{abstract}

\begin{figure}[!t]
DESY--03--85 \\
July 2003
\end{figure}

\section{Introduction} \label{s:introduction}

Measurements of deep-inelastic lepton-proton scattering (DIS) provide
information that is crucial to our understanding of proton structure.  Since
the fixed target experiments have discovered scaling
violations~\cite{Fox:1974ry,deGroot:1978hr}, much progress has been made in
extending the kinematic regime covered in terms of the Bjorken variable $x$ and
the four-momentum transfer $Q^2$.  This holds especially for the HERA $ep$
scattering experiments which, with their wealth of data, have shown that the
$Q^2$ evolution of the proton structure function $F_2(x,Q^2)$ is well described
by perturbative Quantum Chromodynamics (pQCD) throughout a wide range in $x$
and $Q^2$\, \cite{Adloff:2000qk,Aid:1996au,Derrick:1996hn,Adloff:1997yz}.
However, at small $Q^2$ deviations from pQCD predictions are
observed~\cite{Adloff:1997yz,Breitweg:2000yn}, indicating the transition into a
regime where non-perturbative effects dominate and the data can only be
described by phenomenological models such as those derived from the Regge
approach~\cite{Collins:1977jy}.

In order to study this non-perturbative regime, the structure function $F_2$
has been measured at very low values of $Q^2$ and $x$, which are accessible at
HERA via special devices mounted close to the outgoing electron beam
direction~\cite{Breitweg:2000yn} thus facilitating measurements of the
scattered electron at very low angles. These devices, however, do not cover the
transition region at \mbox{$Q^2$ $\sim 1\;$GeV$^{2}$}, which up to now has only
been investigated using ``shifted vertex''
data~\cite{Derrick:1996ef,Adloff:1997mf}. In this paper the possibility to
extend the kinematic domain of HERA into this region using QED Compton (QEDC)
events, {\itshape i.e.} $ep$ events with wide angle hard photon radiation, is
discussed.

The present studies are based on a modified version~\cite{Lendermann:2002} of
the COMPTON event generator~\cite{Carli:1991yn} with a complete description of
the low mass hadronic final states based on the SOPHIA
model~\cite{Mucke:2000yb}, as described in section~\ref{s:simulation}. Apart
from studying the potential of QEDC events to access the low $Q^2$ region this
new version of the COMPTON program thus also allows investigating the
possibility to measure in the fixed target region at higher $x$, because the
result of such a measurement crucially depends on the accurate description of
the hadronic final state at low masses $W$.

\begin{figure}[tb]
\centerline{\epsfig{figure=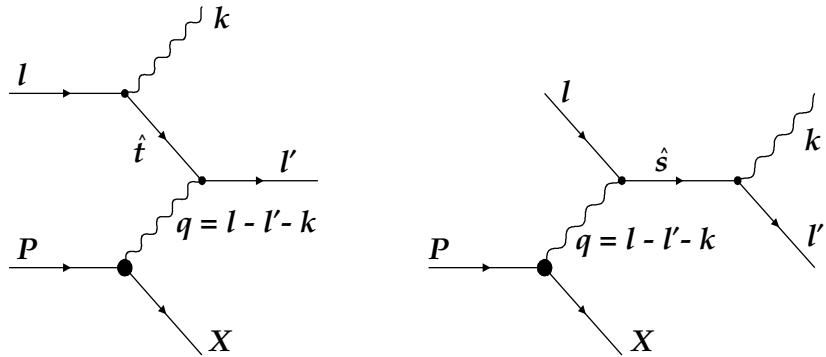,width=.7\textwidth}}
\caption{Lowest order Feynman diagrams for the radiative process $ep 
\rightarrow e{\gamma}X$ with photon emission from the electron line.
$l$, $P$ represent the four-momenta of the incoming electron and the 
incoming proton, while $l^{\prime}$, $k$ and $X$ are the momenta of 
the scattered electron, the radiated photon and the hadronic final 
state, respectively. $\hat{s}$ and $\hat{t}$ denote the squared
four-momenta of the virtual lepton.} 
\label{f:feynman_rad}
\end{figure}

The possibility to measure the proton structure function $F_2$ at low $Q^2$
using QEDC events was first discussed by Bl\"umlein {\itshape et
al.}~\cite{Blumlein:1991kz,Blumlein:1993ef}. In the framework of the equivalent
photon approximation the authors introduced the concept of a photon density of
the proton\footnote{Sometimes also denoted as $f_{\gamma|p}$ or
$D_{\gamma|p}$.}, $\gamma$, to be valid at very low virtualities $Q^2$. This
function has been computed by de R\'ujula and Vogelsang~\cite{DeRujula:1999yq}
when proposing an extraction method for $\gamma$ from HERA QEDC data. The
validity of this approach is discussed in the second part of the paper, where
it is shown that the collinear approximation does not provide a sufficient
description of inelastic QEDC scattering and that measurements of the QEDC
cross section with reasonable precision can thus not be interpreted in terms of
the photon density function.

\section{QEDC Monte Carlo Simulation} \label{s:simulation}

Radiative processes in $ep$ scattering, as depicted in
Fig.\,\ref{f:feynman_rad}, may be split into three different
classes~\cite{Courau:1992ht,Ahmed:1995cf} with (i) the bremsstrahlung or
Bethe-Heitler process corresponding to small masses of both the virtual
electron and the virtual photon, (ii) the QED Compton process with a low
virtual photon and a large virtual electron mass and finally (iii) the
radiative DIS process where the photon is collinear either with the incoming
(Initial State Radiation, ISR) or the outgoing (Final State Radiation, FSR)
electron.  All three classes correspond to distinct experimental signatures. 
For the QEDC scattering process the final state topology is given by an
azimuthal back-to-back configuration of the outgoing electron and photon
detected under rather large scattering angles. In this configuration their
transverse momenta balance such that very low values of the exchanged photon
virtuality $Q^2$ are experimentally accessible.

\begin{figure}[tb]
\centerline{\epsfig{figure=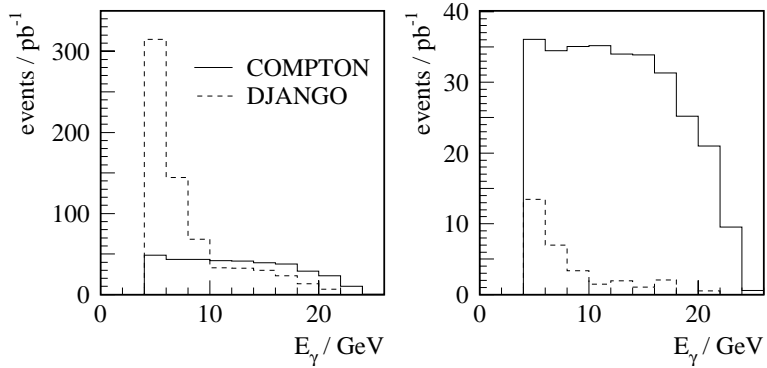,width=.65\textwidth}}
\caption{Energy of the photon candidate in the backward calorimeter
as calculated by the H1 simulation program. Shown are the expectations from
QEDC scattering as predicted by the COMPTON generator
in comparison to the DIS background simulated using DJANGO.
The left and right plots show the distributions before and after applying
the cut of 90$^\circ$ on the maximum polar angle of any further energy
deposition in the calorimeters (see text).}
\label{f:ephoton}
\end{figure}

To correctly describe the process $ep\rightarrow e\gamma X$ the standard
kinematic variables $x$ and $Q^2$, used to describe inclusive deep-inelastic
scattering (DIS), have to be redefined in order to account for the additional
photon in the final state
\begin{equation}
{Q}^{2}=-q^{2}=-(l-l^{\prime}-k)^{2} \;\;, \;\;\;
{x}=\frac{{Q}^{2}}{2P\cdot(l-l^{\prime}-k)} \;\;.
\label{eq=kin_inv}
\end{equation}
Here $l$ and $P$ are the four-momenta of the incoming electron and the incoming
proton, and $l^{\prime}$ and $k$ represent the momenta of the scattered
electron and the radiated photon, respectively (Fig.\,\ref{f:feynman_rad}). 
Three further independent variables are needed for a full description of the
differential QEDC scattering cross section. In the formalism presented
in~\cite{Courau:1992ht} the Lorentz invariant scale variable $x_{\gamma}=q\cdot
l/P\cdot l$ and the scattering solid angle $\Omega^{*}$ defined in the
centre-of-mass frame of the virtual Compton process and encapsulating two
degrees of freedom are employed. The cross section is then given
by~\cite{Courau:1992ht}
\begin{equation}
    \label{e:QEDCxsec}
    \frac{d^{4}\sigma^{ep\rightarrow e\gamma
    X}}{d{x}dx_{\gamma}d{Q}^{2}d\Omega^{*}} =
    f^{T}_{\gamma^{*}/p}({x},x_{\gamma},{Q}^{2}) \left[
    \frac{d\sigma}{d\Omega^{*}} \right]^{T} +
    f^{L}_{\gamma^{*}/p}({x},x_{\gamma},{Q}^{2}) \left[
    \frac{d\sigma}{d\Omega^{*}} \right]^{L} \;\;, 
\end{equation}
where $[ d\sigma/d\Omega^{*} ]^{T,L}$ are the differential cross sections of
the process $e\gamma^{*} \rightarrow e\gamma$ for transverse and longitudinal
polarised photons, fully calculable in the framework of
QED~\cite{Courau:1992ht}, and $f_{\gamma^{*}/p}^{T,L}$ represent the
corresponding virtual photon spectra, which may be expressed in terms of the
photo-absorption cross sections $\sigma_{\gamma^{*}p}^{T,L}$. In order to
specify $\sigma_{\gamma^{*}p}^{T,L}$ one has to consider three separate
contributions depending on the value of the invariant mass $W$ of the outgoing
hadronic final state:
\begin{enumerate}
    \item {\rm Elastic scattering}, for which the proton stays intact
    ($W = m_{p}$).  This channel is well measured, and the cross
    section is given by the electric and magnetic form factors $G_E$ and~$G_M$;
    \item {\rm Resonance production}, where the total mass of the
    hadronic final state $X$ lies in the range $m_p + m_{\pi}
    \lesssim W \lesssim 2$\,GeV;
    \item {\rm Continuum inelastic scattering} at $W \gtrsim
    2$\,GeV. In this region the $\gamma^{*}p$ cross section is
    defined through the proton structure functions $F_2$ and $F_L$.
\end{enumerate}
The above cross section expression~(\ref{e:QEDCxsec}) is implemented in the
COMPTON event generator~\cite{Carli:1991yn}.  However, as this generator was
primarily written for an application in analyses of elastic QEDC events, in the
original version a rather crude approach has been employed to describe the
resonance region  and only simple scale invariant $F_2$ parameterisations are
used to model the continuum inelastic domain. Furthermore, no hadronisation of
the final state $X$ is performed.

As this paper aims at the investigation of inelastic QEDC events a new version
of the COMPTON generator was developed~\cite{Lendermann:2002} which includes
detailed parameterisations for the resonance~\cite{Brasse:1976bf} and the
continuum~\cite{Abramowicz:1997ms} regions. In addition, several packages for a
complete simulation of the hadronic final state have been implemented into the
program. For the present studies the SOPHIA package~\cite{Mucke:2000yb} is used
in the range of low $Q^2$  and low masses, $W$, of the hadronic final state
while the Quark Parton Model with subsequent Lund string
fragmentation~\cite{Sjostrand:2000wi} is employed at high $W$ and high $Q^2$.

\section{On the Measurement of $F_2$ Using QEDC Scattering} \label{s:f2}

The presented studies are based on a Monte Carlo sample which was generated
using the new version of the COMPTON program and corresponds to an integrated
luminosity of about 30~pb$^{-1}$; the incident beam energies used are $E_e =
27.6$\;GeV for the electron and $E_p = 820$\;GeV for the proton beam. In order
to obtain a realistic simulation of the experimental conditions at the HERA
$ep$ detectors, the generated events were subject to the GEANT-based simulation
of the H1 detector~\cite{Abt:1997hi}. Hence, the selection criteria are adapted
to the resolution and acceptance limits of the detector components relevant for
this analysis. The most important ones are the backward
calorimeter\footnote{The $z$ axis of the right-handed coordinate system used by
H1 is defined to lie along the direction of the incident proton beam and the
origin to be at the nominal $ep$ interaction vertex; the backward direction is
thus defined through $z<0$.} SpaCal~\cite{Appuhn:1996vf}, the liquid-argon
(LAr) calorimeter and the tracking and vertex
detectors~\cite{Abt:1997hi,Arkadov:2000}.

\begin{table}[t]
\tablecaption{Summary of QEDC selection criteria as described in text.}
\begin{center}
\begin{tabular}{l@{\hspace{2cm}}l}
  \multicolumn{1}{l}{\bfseries Item} & \multicolumn{1}{l}{{\bfseries Cut value}} \\ 
  \hline 
  QEDC signature          & \rule[-2mm]{0mm}{6mm} 
                              $E_{\gamma}, E_{e}>4$\;GeV                    \\ 
                          & \rule[-2mm]{0mm}{6mm} $E_e+E_{\gamma}>20$\;GeV  \\
                          & \rule[-2mm]{0mm}{6mm}
                              $A=180^{\circ}-\Delta\phi < 45^{\circ}$       \\
  Hadronic final          & \rule[-2mm]{0mm}{6mm} 
                              $E > 0.5$\;GeV (within one cluster)   \\
 \raisebox{.2cm}{\,state} & \rule[-2mm]{0mm}{6mm} 
                              $\theta < 90^{\circ}$  \\ 
  Event properties        & \rule[-2mm]{0mm}{6mm} 
                              existence of reconstructed vertex             \\
  \hline
\end{tabular} 
\label{t:selection}
\end{center}
\end{table}

The analysis strategy is based on the detection of the outgoing photon and the
outgoing electron observed almost back-to-back in azimuth. Thus, the main
experimental requirement is the existence of two electromagnetic energy
depositions (clusters) with $E_{\gamma}, E_{e}>4$\;GeV in the backward
calorimeter; an additional limit is imposed on the total $e\gamma$ energy
$E_{e}+E_{\gamma} > 20$\;GeV to reduce radiative corrections.  In order to
separate QED Compton scattering from the FSR process and still obtain inelastic
QEDC events the azimuthal acoplanarity angle $A=180^{\circ}-\Delta\phi\,$ is
required to be below $45^{\circ}$, where $\Delta\phi$ represents the angle
between the transverse momenta of the electron and the photon; this cut also
matches a corresponding requirement in the COMPTON event
generator~\cite{Courau:1992ht}. Inelastic QEDC events are selected demanding at
least one LAr cluster with energy above $E > 0.5$\;GeV; as the acceptance of
the LAr calorimeter is limited to $\theta \gtrsim 4^{\circ}$ this cut --- apart
from rejecting elastic events --- substantially reduces the inelastic
contribution at very low masses $W$.  After applying these criteria, background
contributions from standard DIS events only appear for large angles of the
hadronic final state particles where one hadron fakes a photon signal in the
backward calorimeter and the current jet is observed in the backward region of
the detector. This contribution is reduced by requiring no additional energy
deposition apart from the two clusters to be found above $\theta_{\rm max} =
90^{\circ}$ (see Fig.\,\ref{f:ephoton}).  Finally, only events with a
reconstructed vertex are considered, in order to guarantee a correct
determination of all kinematic variables. However, as the final state hadrons
are generally scattered under very small angles the corresponding tracks do not
provide sufficient resolution for an accurate vertex measurement. Vertices are
thus reconstructed using the electron track detected in one of the tracking
detectors. A brief summary of all selection criteria is given in
Tab.~\ref{t:selection}.

\begin{figure}[!tb]
\begin{minipage}{.63\textwidth}
\centerline{\epsfig{figure=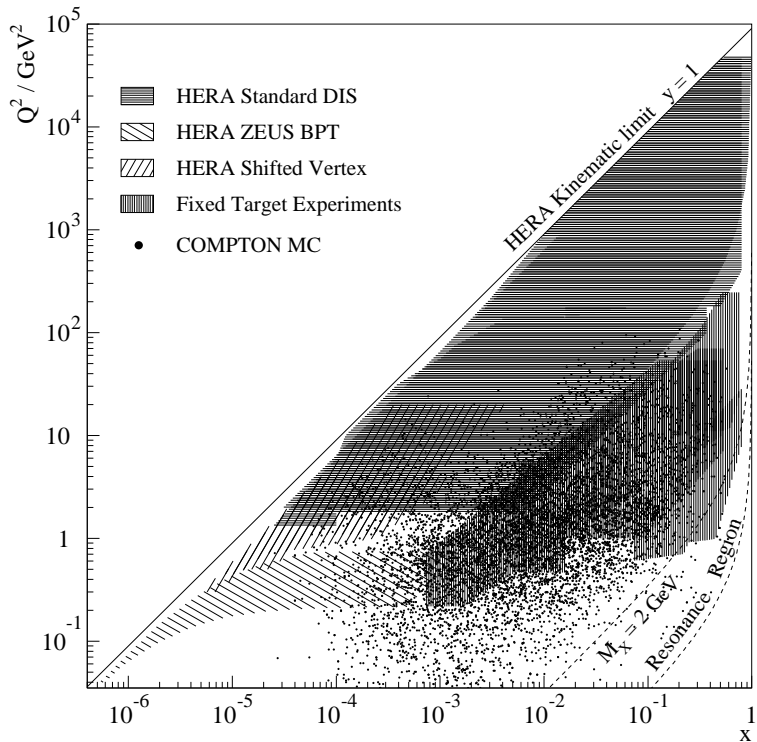,width=\textwidth}}
\end{minipage}\hfill
\begin{minipage}{.31\textwidth}\vspace{4cm}
\caption{\label{f:kinplane_com}Kinematic domain of continuum inelastic
QED Compton events in 
comparison to the regions covered by inclusive DIS measurements at HERA 
and fixed target experiments.}
\end{minipage}
\end{figure}

The phase space covered by a QEDC sample selected by the cuts described is
shown in Fig.\,\ref{f:kinplane_com} for an integrated luminosity of
30\;pb$^{-1}$. Compared to the kinematic range accessed at HERA via standard
deep-inelastic scattering, the QEDC events clearly extend to lower $Q^2$.  For
inclusive DIS the outgoing electron is not detected for such low values of
$Q^2$ as it is scattered at small angles escaping through the beam pipe unseen;
QEDC events, however, with the electron and photon in the final state balancing
in transverse momentum, reach into the transition region below
$Q^2<1.5$\;GeV$^{2}$, which otherwise is only accessed through data taken in
the shifted vertex runs~\cite{Derrick:1996ef,Adloff:1997mf} or with dedicated
devices for tagging events at very low $Q^2$ like the ZEUS
BPT~\cite{Breitweg:2000yn}. But, due to acceptance limitations in the high~$x$
domain, these latter data cannot extend the low $Q^2$ $F_2$-measurements to
such high $x$ as QEDC events. It is therefore the range of medium to high $x$
which is of special interest when analysing QEDC scattering. As higher $x$
correspond to low masses $W$, special emphasis had to be put into the correct
modelling of the hadronic final state. 

\begin{figure}[tb]
\centerline{\epsfig{figure=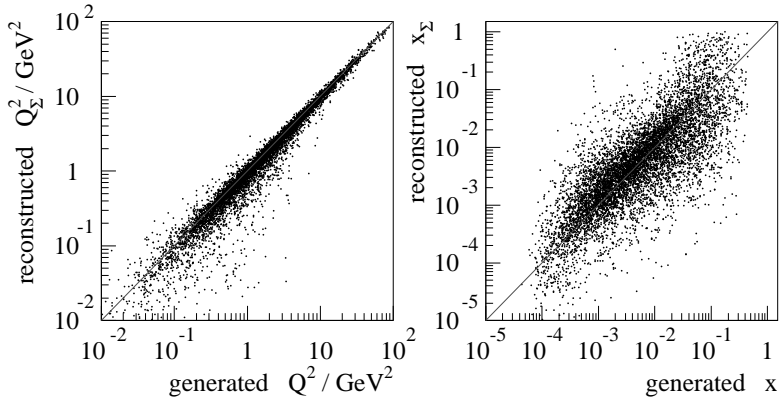,width=.65\textwidth}}
\caption{Correlation between the generated and reconstructed variables 
$Q^2$ and $x$ using the Sigma method as described in the text.}
\label{f:reconstruction}
\end{figure}

An accurate description of the hadronic final state is especially important as,
for the kinematic range in question, the reconstruction of the Bjorken-variable
$x$ cannot be performed using the kinematics of the outgoing electron and
photon; the $x$-resolution of this method deteriorates with $1/y=xs/Q^2$ thus
becoming inapplicable at low values of the inelasticity $y$.  For a double
differential measurement of the structure function $F_2(x,Q^2)$ at $Q^2 \sim
1$\;GeV and $x \gtrsim 10^{-4}$ the variable $x$ has thus to be reconstructed
from the final state hadrons.  This is done using the so-called Sigma
method~\cite{Bassler:1995uq}, which is based on the measurement of $\sum_{i} (E
- p_{z})_{i}$ summing over all objects $i$ of the hadronic final state.  To
suppress the influence of calorimeter noise substantially contributing at low
masses $W$, we use a simple approach in which only clusters with energies above
the noise level of $0.5$\;GeV are considered when reconstructing $x$.
Fig.\,\ref{f:reconstruction} shows the correlation between the generated and
reconstructed values of $Q^2$ and $x$.  It demonstrates the possibility to
reconstruct the QEDC event kinematics using the hadronic final state throughout
the relevant phase space region, {\itshape i.e.} for $0.1 \lesssim Q^2 \lesssim
10$\;GeV$^{2}$ and $10^{-4} \lesssim x \lesssim 10^{-1}$.

\begin{figure}[tb]
\centerline{\epsfig{figure=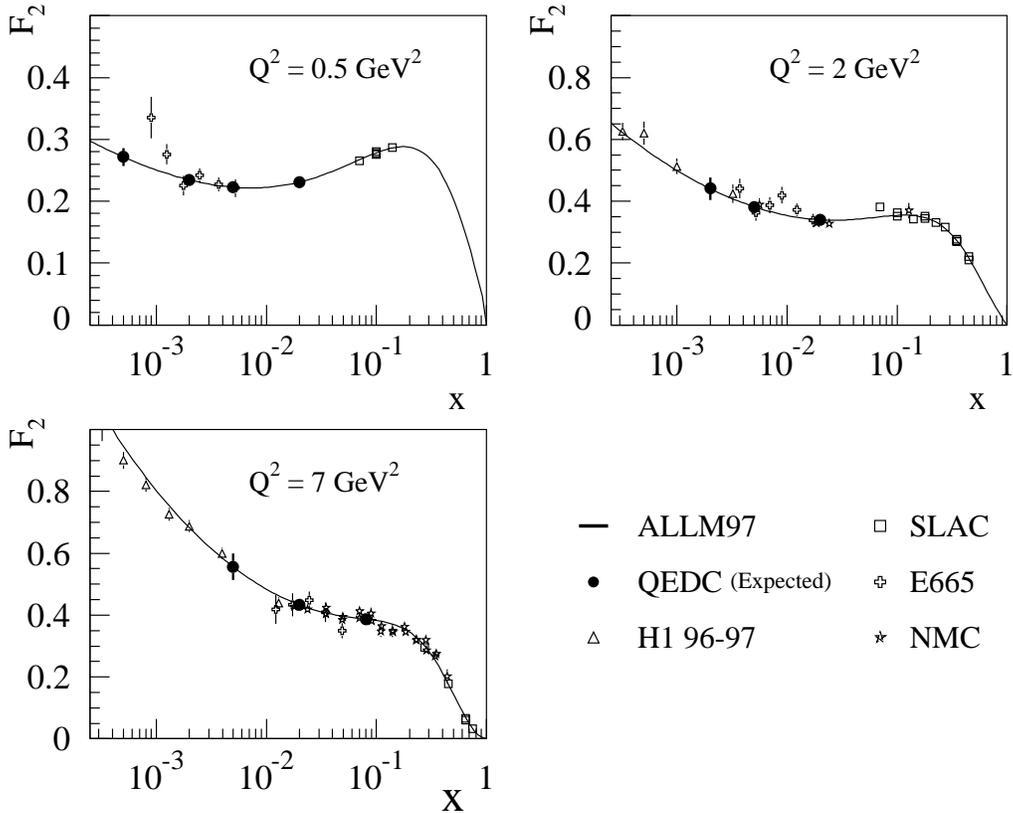,width=.85\textwidth}}
\caption{Possible $F_2$ measurement as expected from a 30\;pb$^{-1}$ QEDC 
event sample. The expected HERA results (closed circles), estimated using the 
COMPTON Monte Carlo program, 
are shown in comparison to H1 results obtained from standard DIS data 
(open triangles~\cite{Adloff:2000qk}) taken during the years 1996-1997 
and measurements from 
fixed target experiments (open squares~\cite{Whitlow:1992uw}, open 
stars~\cite{Arneodo:1997qe} and open crosses~\cite{Adams:1996gu}).}
\label{f:f2meas}
\end{figure}

The expected statistical significance of an $F_2$ measurement based on the same
COMPTON event sample with $30$\;pb$^{-1}$ total integrated luminosity is
presented in Fig.\,\ref{f:f2meas}. In order to extract the structure function
$F_2$ in terms of $x$ and $Q^2$ the selected Monte Carlo data are divided into
subsamples corresponding to a grid in $x$ and $Q^2$.  The bin sizes are adapted
to the resolution in the measured kinematic quantities such that purity and
stability in all bins shown are greater than 30\%; here, the purity (stability)
is defined as the ratio of the number of simulated events originating from and
reconstructed in a specific bin to the number of reconstructed (generated)
events in the same bin.  As the proton structure function $F_2$ is then
obtainted by a bin-by-bin unfolding method using the same Monte Carlo sample,
the extracted $F_2$ values trivially lie on the curve representing the ALLM97
$F_2$ parameterisation used as input to the Monte Carlo generator. Clearly,
the systematic uncertainties remain to be studied in order to judge on the 
overall precision of the measurement.

The results demonstrate that an analysis of QEDC scattering data at HERA would
add information in the low $Q^2$ and medium to high $x$ region not yet covered
by the HERA $ep$ experiments. Such an analysis would extend the present HERA
measurements into the region previously covered only by fixed target data.
 
\section{On the Photon Content of the Proton} \label{s:dics}

In contrast to the exact treatment of the QEDC scattering process, the concept
of the photon content in the proton based on the collinear or equivalent photon
(Weiz\"acker-Williams) approximation~\cite{Weizsacker:1934sx,Williams:1934ad}
provides a much simpler approach to QEDC scattering and is believed to reveal
basic features of photon-induced reactions involving proton beams. In this
``parton model'' approach the transverse component of the exchanged photon
momentum is neglected and the emitted photon is assumed to be on-shell and
collinear with the incident proton. This simplifies the expression for the QEDC
cross section to~\cite{Blumlein:1989gk}
\begin{equation} 
\label{eq:sig_dics2}
        \frac{d^2 \sigma^{\rm ep\rightarrow e\gamma X}}{dx_l\, dQ^2_l} = 
        \frac{2 \pi \alpha^2}{x_l s} 
            \frac{1 + (1 - y_l)^2}{1 - y_l} \ \gamma(x_l,Q^2_l) \;\;\;,
\end{equation}
where the ``structure'' function $\gamma(x_l,Q^2_l)$ parameterises the
photon-parton content of the proton depending only on two degrees of freedom,
the leptonic variables  $Q^2_l = l - l'$ and $x_l = \frac{Q^2_l}{2 P \cdot
q_l}$; at fixed centre-of-mass energy $\sqrt{s}$ the inelasticity $y_l$ is also
defined through these two variables via the relation $y_l = Q^2_l/x_l s$.

According to the above expression, an experimental determination of the double
differential QED Compton scattering cross section $d^2 \sigma / dx_l\, dQ^2_l$
can be interpreted as a measurement of the photon-parton density function
$\gamma(x_l,Q^2_l)$. This $\gamma$ function can then be applied for the
computation of other $ep$ and $pp$ cross sections where the underlying process
is mediated by a quasi-real photon
exchange~\cite{Drees:1994zx,Ohnemus:1994qw,Ohnemus:1994xf,Gluck:1994vy,Gluck:
2002fi,Gluck:2002cm}. The possibility to measure the structure function $\gamma
(x_l,Q^2_l)$ using QEDC events was discussed in several
publications~\cite{Blumlein:1991kz,Blumlein:1993ef,DeRujula:1999yq,Anlauf:
1991vc}. We consider here the latest work by De~R\'ujula and
Vogelsang~\cite{DeRujula:1999yq} where they proposed to compare the $Q^2_l$
dependence of $\gamma$ at fixed $x_l$ with its gluon counterpart $g(x,Q^2)$.

\begin{figure}[!tb]
\centerline{\epsfig{figure=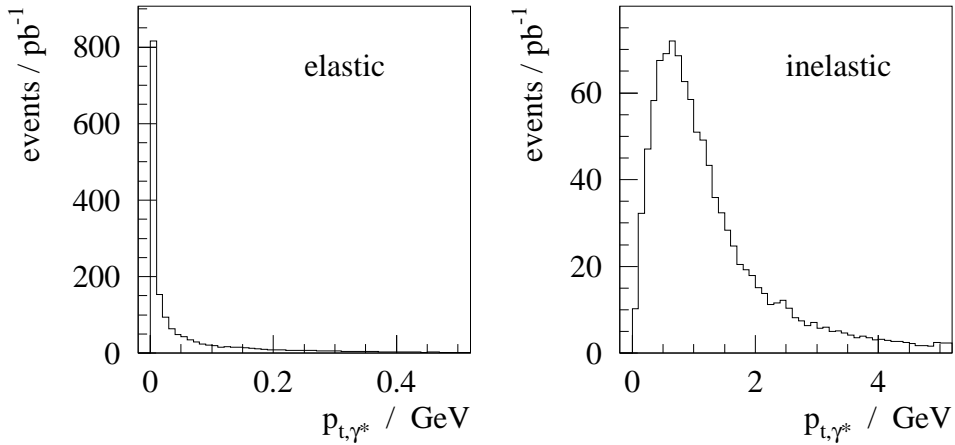,width=0.8\textwidth}}
\caption{Transverse momentum of the exchanged photon in elastic (left)
and inelastic (right) QEDC events generated by the COMPTON program.}
\label{f:collinear}
\end{figure}

In order to judge on the feasibility of measuring $\gamma$ a dedicated study of
the accuracy of the collinear approximation is performed.
Fig.\,\ref{f:collinear} shows the transverse momentum ($p_{t,\gamma^\ast}$)
distribution of the exchanged photon in elastic and inelastic QEDC events as
predicted by the COMPTON event generator. While for elastic events
$p_{t,\gamma^\ast}$ is rather small, much larger values are reached when
selecting inelastic scattering processes. Thus, as in the collinear
approximation one assumes $p_{t,\gamma^\ast}$ to vanish, one can expect
significant deviations when calculating event kinematics. The effect is further
enhanced due to the acceptance constraints, which demand that both outgoing
particles, electron and photon, are measured under finite polar angles. For the
presented studies the QEDC selection criteria listed in Table~\ref{t:selection}
are applied on generated quantities; in addition the detector acceptance is
approximately reproduced by demanding $0.06 < \theta_e, \theta_\gamma < \pi -
0.06$. These criteria are close to those proposed in~\cite{DeRujula:1999yq}.

\begin{figure}[!p]
\centerline{\hbox{\psfig{figure=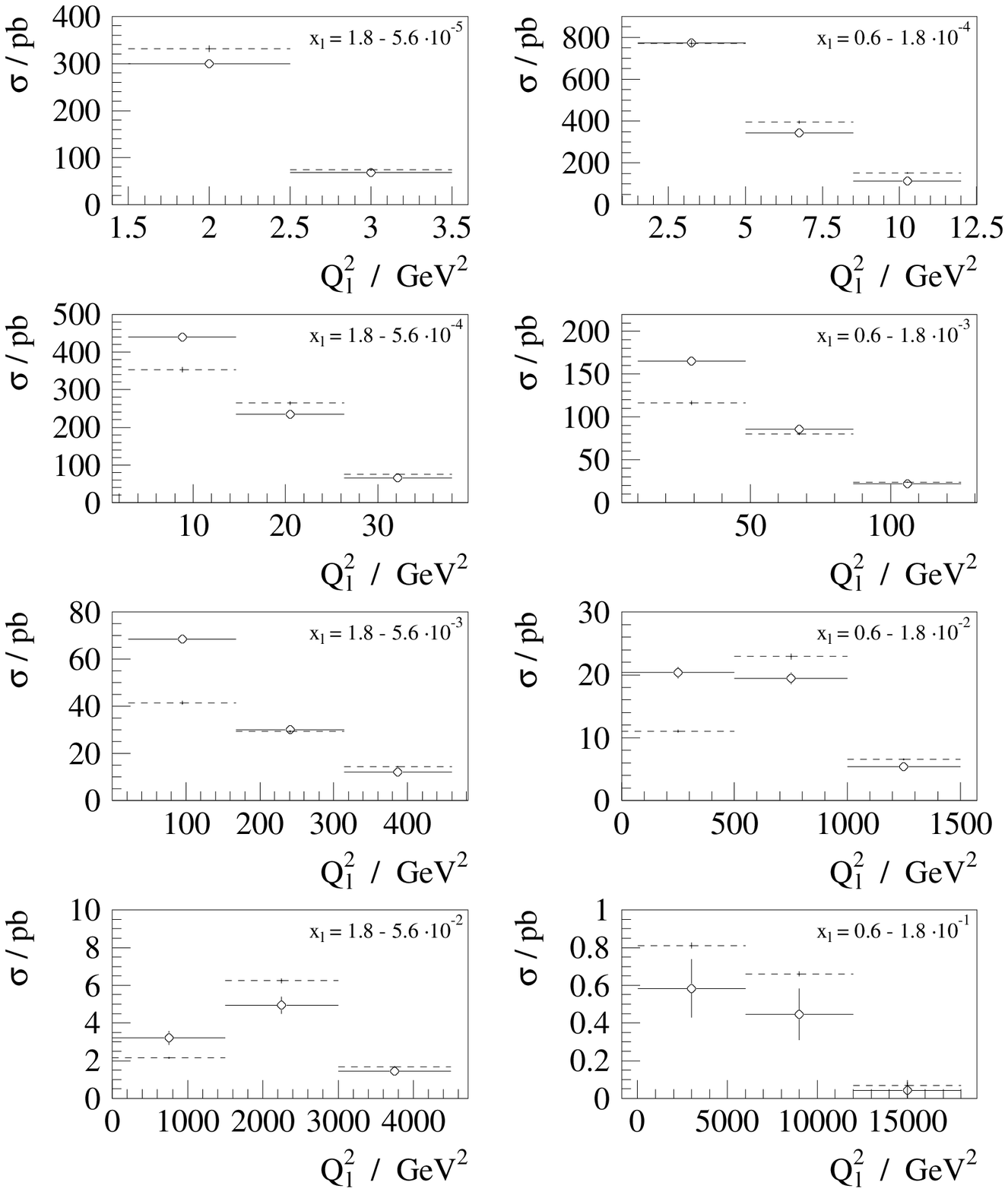,width=0.8\textwidth}}}
\caption{Double differential cross section of the QED Compton scattering.
  Open circles depict cross section values given by the
  COMPTON generator. The corresponding error bars show statistical errors.
  The dashed lines denote the values computed by W.\,Vogelsang.}
\label{f:xsection_tot}
\end{figure}

A comparison between the exact calculation and the predictions from the
collinear approximation is given in Fig.\,\ref{f:xsection_tot}, which shows the
total, i.e. the sum of the elastic and inelastic QEDC cross section as a
function of $Q_l^2$ in bins of $x_l$. Significant discrepancies are observed in
several bins. The exact cross sections were derived from a sample of COMPTON
events generated without radiative correction. For the collinear approximation
predictions have been computed analytically by Vogelsang~\cite{Vogelsang:2001}
using the same set of cuts; the corresponding uncertainty of this calculation
was estimated to be approximately 2\%. Furthermore the elastic scattering cross
sections generated by the COMPTON program and computed in the
Weizs\"acker-Williams approximation were compared. Very good agreement was
obtained showing that the collinear approximation provides, indeed, a good
description of the elastic process.

It has been checked that the difference between the $F_2$ parameterisations
employed in the COMPTON generator and in the calculations by De~R\'ujula and
Vogelsang, cannot account for the observed discrepancies: the different
predictions for $\gamma$ calculated according to the relations given
in~\cite{Blumlein:1993ef,DeRujula:1999yq} agree within 4\%~\footnotemark. The
inelastic QEDC cross sections, however, deviate by up to a factor of~2 from
each other. This shows, that apparently the equivalent photon approximation
does not provide a sufficient degree of accuracy in the inelastic region and
that the factorisation of the cross section given by Eq.~(\ref{eq:sig_dics2})
in terms of only two kinematic variables is a too rough approximation of the
QEDC scattering process.

\footnotetext{As the phenomenological $F_2$ parameterisations used in the
COMPTON program have, contrary to the pQCD based PDFs employed in 
\cite{DeRujula:1999yq}, no artificial lower $Q^2$ limit, we also tested
different start scales $Q^2_0$ for the integration over $F_2$,
as specified in \cite{DeRujula:1999yq} and Eq.~(1) of \cite{Blumlein:1993ef}.
The differences in $\gamma$ are in any case of the order of a few percent.}

\begin{figure}[!p]
\centerline{\hbox{\psfig{figure=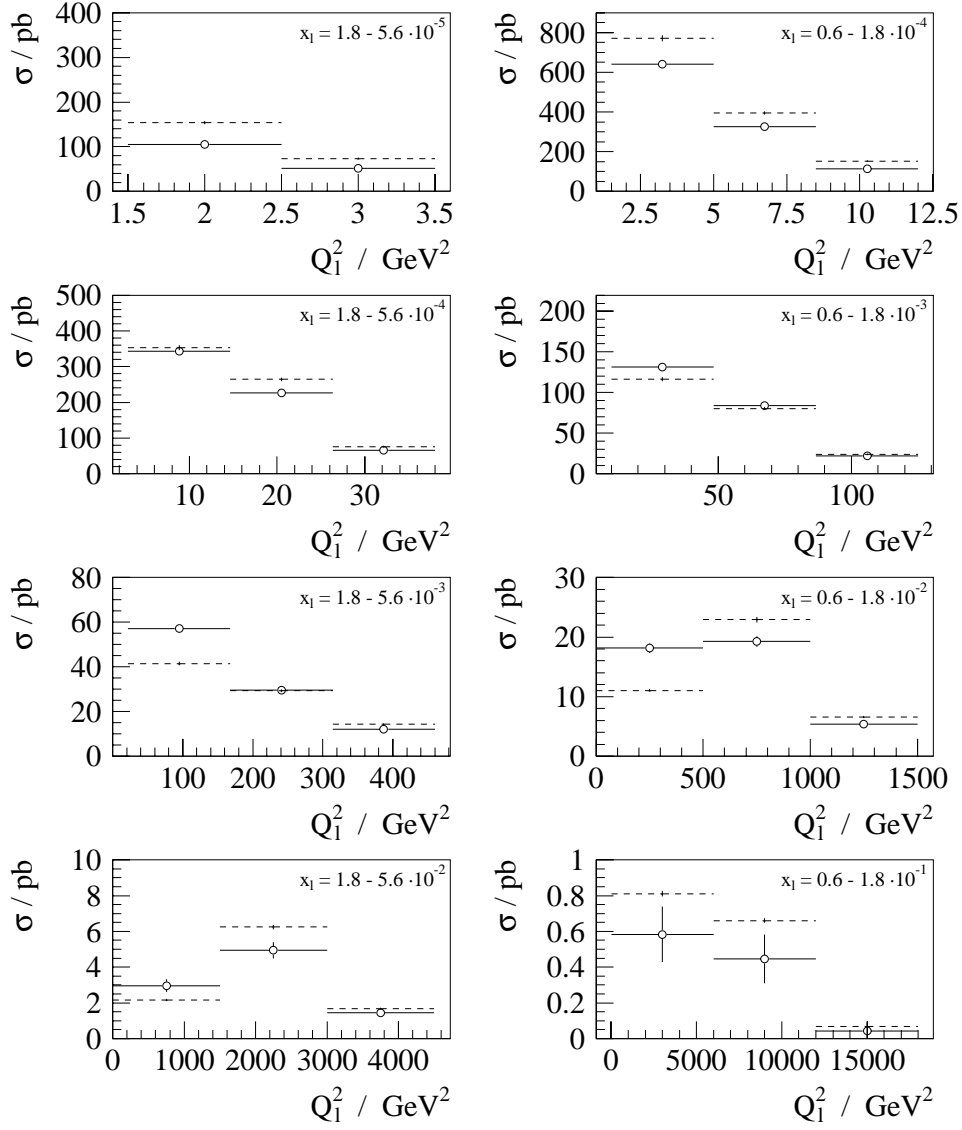,width=0.8\textwidth}}}
\caption{QED Compton scattering cross section after additional
  kinematic cuts. Open circles depict cross section values given by the
  COMPTON generator. The corresponding error bars show statistical errors.
  The dashed lines show the values computed by W.\,Vogelsang.}
\label{f:xsection_drv}
\end{figure}

In order to ensure the validity of the collinear approximation, additional
selection cuts were proposed in~\cite{DeRujula:1999yq}: \begin{equation}
\label{eq:kincuts} -\hat{t}, \hat{s} > 1\,{\rm GeV}^2 \hspace{3em} {\rm and}
\hspace{3em} p_{t,e}, p_{t,\gamma} > 1\,{\rm GeV}\ , \end{equation} where the
momentum scales $\hat{t}$ and $\hat{s}$ represent the virtualities of the
exchanged lepton in the Feynman diagrams shown in Fig.\,\ref{f:feynman_rad}.
Here, the cuts applied on the transverse momenta $p_{t,e}$ and $p_{t,\gamma}$,
actually, have a much stronger effect on the distributions of COMPTON events
than those imposed on $\hat{t}$ and $\hat{s}$. After applying all of the four
additional cuts the COMPTON QEDC cross section is again compared to the
corresponding analytical calculations provided by Vogelsang. As shown in
Fig.\,\ref{f:xsection_drv} there are, as before, significant discrepancies
observed.

Fig.\,\ref{f:xsection_drv} illustrates the difficulties of the collinear
approximation. For $Q^2_l<5$~GeV$^2$ and $5.6 \cdot 10^{-5} < x_l < 1.8 \cdot
10^{-3}$ the cross sections predictions from the COMPTON program drop
significantly when applying the additional cuts on $\hat{t}$, $\hat{s}$ and
$p_{t,e}$, $p_{t,\gamma}$. This is not the case for the equivalent photon
approximation where the cross sections remain unchanged due to an incorrect
calculation of the event kinematics arising from the asumption
$p_{t,\gamma^\ast}=0$.

It is not ruled out that some carefully chosen cuts on the transverse momenta,
polar angles and other event quantities may limit the phase space such that the
equivalent photon approximation becomes applicable also for inelastic QEDC
scattering. However, with such limitations it is questionable whether the
$\gamma$-function describes a characteristic and universal property of the
proton relevant for precision measurements.

\section{Concluding Remarks} \label{s:summary}

The presented analysis shows that QEDC scattering data at HERA provide the
potential to measure the proton structure function $F_2$ at low $Q^2$ and
medium to high $x$, extending the kinematic range covered so far into the fixed
target regime. As such a measurement requires a good understanding of the
hadronic final state at low invariant masses, an improved version of the
COMPTON generator was developed which allows for a better event modelling. With
this generator the accuracy of calculations using the equivalent photon
approximation is studied. It is revealed that this approximation is not able to
describe the inelastic QEDC cross section with sufficient precision. A
measurement of the photon density of the proton from QEDC events would, thus,
not provide any decisive insight into the proton structure.

\section{Acknowledgements} \label{s:acknowledgements}

We would like to thank W.\,Vogelsang for the fruitful cooperation in the study
of the collinear approximation. We are grateful to the H1 collaboration for
providing the H1 simulation and reconstruction programs. This work was
supported by the Federal Ministry for Education, Science, Research and
Technology, FRG, under contract number \mbox{05\,H1\,1PEA/6}.

\end{document}